\documentclass{llncs}

\usepackage{graphicx} % Required for inserting images
\usepackage{amsmath,bm}
\usepackage{mathtools}
\usepackage{amssymb}
\usepackage{xcolor} % colors for editorial markup; does not change layout
\usepackage{url}    % for \url in footnotes
\newcommand{\todo}[2][]{ }

\usepackage{tikz}
\usepackage{float}
\usetikzlibrary{arrows.meta}
\usepackage{booktabs}

\newcommand{\R}{\mathbb{R}}

\newcommand{\calC}{\mathcal{C}}
\newcommand{\calH}{\mathcal{H}}

\DeclareMathOperator{\Colspan}{ColSpan}
\DeclareMathOperator{\Relu}{ReLU}
\DeclareMathOperator{\Rank}{Rank}

\title{Beyond Scalar Probes: Exploiting Vector-Valued Outputs in ReLU Networks For Signature Extraction}

\author{
Gorka Abad\inst{1} \and
Claude Carlet\inst{1,4} \and
Ermes Franch\inst{1} \and
Stjepan Picek\inst{1,2,3} \and
Vincent Rijmen\inst{1,5}
}

\institute{
University of Bergen, Bergen, Norway\\
\email{\{ermes.franch,gorka.abad\}@uib.no}
\and
University of Zagreb, Zagreb, Croatia
\and
Radboud University, Nijmegen, The Netherlands\\
\email{stjepan.picek@ru.nl}
\and
Université Paris 8, Saint-Denis, France\\
\email{Claude.Carlet@univ-paris8.fr}
\and
KU Leuven, Leuven, Belgium\\
\email{vincent.rijmen@esat.kuleuven.be}
}

\begin{document}

\maketitle

\begin{abstract}
We revisit cryptanalytic extraction of ReLU networks from a geometric and algebraic perspective. Rather than restricting attention to a single output component, we study the full vector-valued behavior across adjacent linear regions. This leads to a rank-one characterization of Jacobian differences that recovers the usual row-signature information while also revealing complementary column-side information. 

Our experiments show how this additional structure can be used in extraction and improves numerical estimation under different numerical-precision regimes ($\mathrm{float64}, \mathrm{float32}$ and $\mathrm{float16}$). We extend the analysis beyond the high-precision and output-rounding settings commonly considered in the literature
towards the low-precision settings encountered in many practical settings.
Code is available online\footnote{\url{https://anonymous.4open.science/r/relu-signature-leakage-BE51/README.md}}.
\end{abstract}


\section{Introduction}
\label{sec:introduction}

Machine learning (ML) and, in particular, deep learning (DL) can represent valuable intellectual property. Their development may require considerable computational resources, large and carefully collected datasets, and substantial engineering effort. This makes the confidentiality of their trained parameters an important aspect of machine-learning security. Model extraction attacks~\cite{tramer2016stealing} are a direct threat to this confidentiality. These attacks aim to imitate as faithfully as possible a target model, exploiting black-box access to the same. In this framework, the attacker can query the target model on chosen inputs and observe the outputs without knowing any other information about internal states or parameters.

Initial approaches aimed at task-accuracy extraction using queries to the original model to train a surrogate that will imitate the original model on a similar task or on a similar data distribution~\cite{tramer2016stealing}. A stronger objective is fidelity extraction, where the goal is to recover a model implementing the same function as the target (i.e., for any input, the output of the stolen model will be exactly the same as the original). Cryptanalytic attacks~\cite{carlini2020} may go further and aim at parameter extraction, recovering the underlying parameters up to transformations that leave the realized function unchanged. These approaches leverage the piecewise linear structure of some neural networks~\cite{Milli19,jagielski2020high,carlini2020}.

Following previous work~\cite{carlini2020}, in this work we consider the setting where the attacker can choose queries and have access to the raw, vector-valued output of the network. Previous works analyze the transition through the boundaries of linear cells to recover the weights and the biases of the model. We instead consider the complete affine maps associated with two adjacent linear cells. If two cells differ only in the activation of one neuron, their Jacobians satisfy:
$$
\bm \Delta = \bm J_1 - \bm J_2 = \bm c \bm r^\intercal,
$$
hence $\bm \Delta$ has rank $1$. The rows of this matrix encode the same neuron information used in~\cite{carlini2020} and follow-up works, while the columns add information lost when observing only a single output component. 

Since all nonzero rows of $\bm \Delta$ are scaled copies of the same direction, they provide redundant observations of the neuron signature. Rather than treating these observations independently, we exploit the known rank-one structure of $\bm \Delta$. Given the noisy observation $\bm S = \bm \Delta  + \bm E$, where $\bm E$ represents the error introduced by finite precision arithmetic, its dominant singular component is the best rank-one approximation of $\bm S$ in Frobenius norm. We use the dominant right singular vector of $\bm S$ as an estimate of the common signature direction, combining the information carried by all output components.

We also investigate two refinements intended to improve numerical precision, at the cost of additional queries. The first maximizes the interval used to estimate the Jacobian in each direction, similarly to~\cite{noman2026rounding}, while the second adjusts an initial estimate of the neuron weights by probing along approximate tangent directions in adjacent linear cells. These refinements can improve estimation when the cell-membership test is reliable, but may degrade accuracy at lower precision when the test accepts intervals crossing cell boundaries.

A side advantage of using the full raw output vector instead of a single component is that, when neuron transitions occur in the last hidden layer, the matrix $\bm \Delta$ exposes column signatures of the final weight matrix, giving a leakage channel complementary to the usual forward extraction of row signatures. These leaks do not directly strengthen end-to-end extraction attacks. Its value is mainly structural, clarifying what information the full vector response contains that scalar-output methods discard. Our contributions can be summarized as:

\begin{enumerate}
    \item We show that a simple transition, in which exactly one ReLU neuron changes activation, induces a rank-one Jacobian difference
    $\bm \Delta =\bm c\bm r^\intercal$
    (Eq.~\eqref{eq:Signature_as_rank1}). Its right factor recovers the row-side information exploited by previous extraction attacks, while its left factor exposes complementary column-side information. We characterize when these column signatures correspond to $\bm W^{(\ell)}$ and discuss the limitations of recovering biases and signs from the column side alone.

    \item We use the rank-one redundancy of $\bm \Delta$ to construct an SVD-based row-signature estimator that combines all output components. Since each oracle query already returns the complete raw-output vector, this additional information can be exploited without increasing the number of oracle queries.

    \item We evaluate the resulting estimator against a Carlini-style single-output baseline under float64, float32, and float16 arithmetic. Our experiments show improved parameter recovery under finite precision, particularly in the float32 regime, and identify the cell-membership test as the main limitation of the adaptive refinements.
\end{enumerate}
\section{Preliminaries}
\label{sec:preliminaries}

\subsection{Notation}
We denote by $\R$ the set of real numbers. The vector space of $n$-tuples of real numbers is denoted by $\R^n$. Vectors are written in bold lowercase letters, such as $\bm v \in \R^{n}$, and are interpreted as column vectors unless otherwise stated. Similarly, the transpose $\bm v^\intercal$ will be interpreted as a row vector. For a vector $\bm v$, its $i$-th component is denoted by $v_i$ where $i$ goes from $1$ to $n$.

Matrices with real entries, $m$ rows and $n$ columns, are elements of $\mathbb{R}^{m \times n}$, and are denoted by bold uppercase letters, such as $\bm M$. The vector formed by the entries of the $i$-th row of $\bm M$ is denoted by $\bm m_{i,*}$, and the $j$-th column of $\bm M$ is denoted by $\bm m_{*,j}$. Both are column vectors; hence, the $i$-th row of $\bm M$ is written as $\bm m_{i,*}^\intercal$.
The transpose of a matrix $\bm M \in \R^{m \times n}$ will be indicated by $\bm M^{\intercal} \in \R^{n \times m}$. Given two functions $f(x),g(x)$, we will indicate their composition as $(f \circ g)(x)= f(g(x))$.

\subsection{Deep Neural Networks}

A deep feedforward neural network consists of several layers. Algebraically,
each hidden layer is the composition of an affine map and a component-wise
activation function.

Let $n_0,\ldots,n_\ell$ be positive integers denoting the layer widths. For
$k=1,\ldots,\ell$, let
\[
    A^{(k)} : \R^{n_{k-1}} \rightarrow \R^{n_k},
    \qquad
    A^{(k)}(\bm h)
      = \bm W^{(k)}\bm h+\bm b^{(k)},
\]
where $\bm W^{(k)}\in\R^{n_k\times n_{k-1}}$ is the \textbf{weight matrix}
and $\bm b^{(k)}\in\R^{n_k}$ is the \textbf{bias vector}. We denote the
entries of $\bm W^{(k)}$ by $w_{i,j}^{(k)}$ and those of $\bm b^{(k)}$
by $b_i^{(k)}$.

Let $\bm h^{(0)}=\bm x$ be the network input and let $\bm\sigma$ denote
the component-wise application of an activation function
$\sigma:\R\rightarrow\R$. For $k=1,\ldots,\ell-1$, define the
preactivation vector $\bm z^{(k)}\in\R^{n_k}$ and hidden activation vector
$\bm h^{(k)}\in\R^{n_k}$ by
\begin{equation}
    \bm z^{(k)}
      = \bm W^{(k)}\bm h^{(k-1)}+\bm b^{(k)},
    \qquad
    \bm h^{(k)}
      = \bm\sigma\!\left(\bm z^{(k)}\right).
\end{equation}

When soft labels are exposed, a softmax is typically applied to the final
affine output, while hard-label models return its argmax. In our setting,
we assume access to the raw output before either operation. Consequently,
the last layer is affine, and its raw network output is
\begin{equation}
    F(\bm x)
      = A^{(\ell)}(\bm h^{(\ell-1)})
      = \bm W^{(\ell)}\bm h^{(\ell-1)}+\bm b^{(\ell)}.
\end{equation}

\begin{definition}[$\ell$-layer deep neural network]
\label{def:DNN}
An \textbf{$\ell$-layer deep neural network} is a function
$F:\R^{n_0}\rightarrow\R^{n_\ell}$ of the form
\begin{equation}
\label{eq:DNN}
    F(\bm x)
    =
    A^{(\ell)}
    \circ \bm\sigma
    \circ A^{(\ell-1)}
    \circ \cdots
    \circ \bm\sigma
    \circ A^{(1)}(\bm x).
\end{equation}
The input layer is not counted among the $\ell$ layers.
\end{definition}

From now on, we restrict to the ReLU activation
\[
    \sigma(x)=\Relu(x)=\max\{0,x\}.
\]

For $k<\ell$, let $\bm w_{i,*}^{(k)\intercal}$ denote the $i$-th row of
$\bm W^{(k)}$. We denote by $\eta_i^{(k)}:\R^{n_{k-1}}\rightarrow\R$
the $i$-th neuron of layer $k$, defined by
\[
    \eta_i^{(k)}(\bm h)
    =
    \Relu\!\left(
        \bm w_{i,*}^{(k)\intercal}\bm h+b_i^{(k)}
    \right).
\]
For a network input $\bm x$, the input to this neuron is
$\bm h^{(k-1)}(\bm x)$.

\begin{definition}[Critical point {\cite{carlini2020}}]
\label{def:critical_point}
The neuron $\eta_i^{(k)}$ is said to be in an \textbf{active state},
\textbf{inactive state}, or \textbf{critical state} at
$\bm h\in\R^{n_{k-1}}$ if
\[
    \bm w_{i,*}^{(k)\intercal}\bm h+b_i^{(k)}
\]
is positive, negative, or zero, respectively.

A point $\bm h\in\R^{n_{k-1}}$ satisfying
\begin{equation}
\label{eq:critical_point}
    \bm w_{i,*}^{(k)\intercal}\bm h+b_i^{(k)}=0
\end{equation}
    
is called a \textbf{critical point}. The set of all critical points is
the \textbf{critical hyperplane}
\[
    \calH_i^{(k)}
    :=
    \left\{
        \bm h\in\R^{n_{k-1}}
        \mid
        \bm w_{i,*}^{(k)\intercal}\bm h=-b_i^{(k)}
    \right\}.
\]
\end{definition}

The critical hyperplane is parallel to the $(n_{k-1}-1)$-dimensional space $\bm w_{i,*}^{(k)\perp} := \{\bm h \in \R^{n_{k-1}} \mid \bm w_{i,*}^{(k)\intercal} \bm h = 0 \}$. In particular, the vector $\bm w_{i,*}^{(k)}$ is orthogonal to the underlying space of the hyperplane $\calH_{i}^{(k)}.$
This observation was exploited in~\cite{shamir_carlini24} to recover the row vectors $\bm w_{i,*}^{(k) \intercal}$ up to some scalar.
An equivalent way of extracting these vectors was previously developed in~\cite{carlini2020} through differential techniques.

\subsection{Invariant Transformation of Neural Network}

In Definition \ref{def:DNN}, we defined $F$ as a composition of affine functions $A^{(k)}(\bm x) = \bm W^{(k)} \bm x + \bm b^{(k)}$ interleaved by activation functions $\bm \sigma.$
Since $\bm \sigma$ acts component-wise, for any permutation matrix
$\bm P\in\R^{n_k\times n_k}$ we have
$\bm \sigma(\bm P\bm h)=\bm P\bm \sigma(\bm h)$.
Therefore, for every hidden layer $k<\ell$, the function $F$ is preserved under the transformation
$$
    F(\bm x) = A^{(\ell)} \circ \cdots \circ (A^{(k+1)} \circ \bm P^{-1} )\circ \bm \sigma \circ (\bm P \circ A^{(k)})\circ  \cdots \circ A^{(1)}(\bm x).
$$
When the activation function is $\Relu$, there is another transformation of the affine functions that preserves $F$.
Indeed, $\Relu(a x)=a\Relu(x)$ for every $a>0$.
More generally, let $k<\ell$ and let
$\bm D\in\R^{n_k\times n_k}$ be a diagonal matrix with strictly positive
diagonal entries. Since ReLU acts component-wise,
\[
    \bm\sigma(\bm D\bm h)=\bm D\bm\sigma(\bm h).
\]
Therefore, the function represented by the network is unchanged when the
parameters of two consecutive layers are transformed as
\[
    \widetilde{\bm W}^{(k)}=\bm D\bm W^{(k)},
    \qquad
    \widetilde{\bm b}^{(k)}=\bm D\bm b^{(k)},
    \qquad
    \widetilde{\bm W}^{(k+1)}=\bm W^{(k+1)}\bm D^{-1}.
\]
Equivalently, the rows of $\bm W^{(k)}$ and the corresponding entries of
$\bm b^{(k)}$ may be scaled by arbitrary positive factors, provided that
the corresponding columns of $\bm W^{(k+1)}$ are scaled by their reciprocals.
This transformation does not apply to the output layer, because there is no
subsequent layer in which to compensate for the rescaling.

\subsection{Signatures}

Row signatures are defined in~\cite{canales23} as follows:
\begin{definition}[Row Signature \protect{\cite[Definition~4]{canales23}}]\label{def:signature}
Let $\eta_i^{(k)}(\bm h) = \sigma(\bm w_{i,*}^{(k) \intercal} \bm h  + b_i^{(k)})$ be a neuron on the $k$-th layer.
The \textbf{row signature}  of $\eta_i^{(k)}$ is the row vector
$$
   (w_{i,1}^{(k)})^{-1} \bm w_{i,*}^{(k) \intercal} = \left (1,\frac{ w_{i,2}^{(k)}}{ w_{i,1}^{(k)}},\ldots, \frac{ w_{i,n_{k-1}}^{(k)}}{ w_{i,1}^{(k)}}\right ).
$$
That is, the $i$-th row of $\bm W^{(k)}$ is normalized on its first component.
\end{definition}
Notice this normalization requires the first component $w_{i,1}^{(k)}$ to be nonzero.
Since the set of such vectors has measure zero (or, rather, a small measure, because of finite precision) and it is extremely unlikely to happen, there is no need to address this particular issue.
In the same way, one can define a \textbf{column signature} for the columns of $\bm W^{(k)}$ normalized on their first components.

In this paper, we will use a slightly different definition of row signature:
\begin{definition}[Row Signature (this work)]\label{def:oursignature}
Let $\eta_i^{(k)}(\bm h) = \sigma(\bm w_{i,*}^{(k) \intercal} \bm h  + b_i^{(k)})$ be a neuron on the $k$-th layer, and assume that $\bm w_{i,*}^{(k)}\neq\bm 0$.
The \textbf{row signature}  of $\eta_i^{(k)}$ is the row vector
$$
    \frac{\mathrm{sign}(w^{(k)}_{i,t})}{\lVert\bm w_{i,*}^{(k)}\rVert_2} \bm w_{i,*}^{(k)\intercal},
$$
where $t=\min\{r:w^{(k)}_{i,r}\neq 0\}$ and $\mathrm{sign}$ is the function outputting $+1$ if the input is positive and $-1$ if it is negative.
That is, the $i$-th row of $\bm W^{(k)}$ is scaled to have norm $1$ whose first nonzero coordinate is positive.
\end{definition}

In a similar way, for a nonzero column $\bm w_{*,j}^{(k)}$, its
\textbf{column signature} can be defined as the vector
\[
    \frac{\mathrm{sign}(w^{(k)}_{t,j})}
         {\lVert\bm w_{*,j}^{(k)}\rVert_2}
    \bm w_{*,j}^{(k)},
    \qquad
    t=\min\{r:w^{(k)}_{r,j}\neq 0\}.
\]
One advantage of the modified definitions is that they also work for rows or columns starting with a zero element.
Additionally, their computation does not incur numerical issues when the rows or columns start with elements close to zero.

Using the invariants described in the previous section, it is possible to construct different algebraic descriptions
of the same neural network. Consequently, we may use positive rescaling to
normalize either the columns of $\bm W^{(2)},\dots,\bm W^{(\ell)}$ or the
rows of $\bm W^{(1)},\dots,\bm W^{(\ell-1)}$. Positive rescaling does not
reverse orientation. Therefore, after normalization, each row or column is
equal to either its canonical signature or the negative of that signature.

The signature of a row determines a unit normal direction for the associated
critical hyperplane, but it does not determine the offset of that hyperplane;
the offset also depends on the corresponding normalized bias. Moreover, even
when the hyperplane is known, the canonical signature does not determine which
side is active, because reversing both the weight vector and the bias preserves
the hyperplane while exchanging its two sides. Recovering this orientation sign
is therefore necessary for functional reconstruction. For a column signature,
the missing sign analogously determines the direction of the neuron's
contribution to the next layer.

\section{Related Work} 
\label{sec:related}

Query-based model reverse engineering predates modern neural-network extraction. Lowd and Meek~\cite{lowd2005adversarial} studied adversarial query strategies for reverse engineering linear and Boolean classifiers.

In 2016, Tramèr et al. pioneered the first attack methods based on synthetic data generation (i.e., interrogating the model on chosen inputs to generate pairs $(x, f(x))$ that can be used to train a surrogate model) and simple query techniques. More precisely, early model extraction attacks typically relied on synthetic data generation or simple interrogation methods to mimic the model's behavior~\cite{tramer2016stealing}. However, these methods have proved to be insufficient to accurately replicate the actual model parameters. Jagielski et al.~\cite{jagielski2020high} followed a different approach that made significant progress towards the extraction goal with high accuracy and high fidelity, although their methods still lacked the query efficiency offered by subsequent cryptanalytic techniques. On the theoretical side, Fefferman~\cite{fefferman1994reconstructing} showed that, under suitable assumptions and with exact knowledge of the realized function, neural-network parameters can be reconstructed. In an orthogonal line of work, Batina et al.~\cite{236204} used electromagnetic side-channel leakage to recover neural-network architecture and parameters.

The attacks closest to ours exploit the piecewise linear structure of $\Relu$ networks.
Milli et al.~\cite{Milli19} used gradient information exposed by model explanations to reconstruct network parameters, while Jagielski et al.~\cite{jagielski2020high} developed high-fidelity extraction from standard model outputs. A significant milestone in this field came with the 2020 paper by Carlini et al.~\cite{carlini2020}, where the problem was framed as a cryptanalytic problem and showed how critical points and derivative discontinuities can reveal neuron signatures with high precision. Compared to~\cite{jagielski2020high}, the attack in~\cite{carlini2020} achieved 220 times more accurate results with 100 times fewer queries.
However, the difficulty of identifying the signs (positive or negative) of neurons in deep networks and the need for an excessive number of input-output pairs to obtain a highly accurate model limited the practical applicability of the attack.

Canales-Mart{\'i}nez et al. developed techniques, among which one they named neuron wiggling~\cite{10.1007/978-3-031-58734-4_1} to solve the problem of identifying the signs of hidden neurons in deep networks and thus reduce the time complexity of the attack to a polynomial level. It is important to note that both attacks mentioned above assume the attacker has access to raw output. 

Hard label extraction was later studied by Chen et al., who proposed an extraction method that requires a polynomial number of queries but suffers from exponential runtime~\cite{chen2024hard}. This theoretical challenge was subsequently overcome by Carlini et al.~\cite{10.1007/978-3-031-91107-1_13}. The authors demonstrated that parameters could be extracted by analyzing the bending of decision boundaries (i.e., points at which the model output switches from one class to another) in proximity to a neuron switch, thereby achieving polynomial-time complexity and making hard-label attacks practically feasible. 
More recently, Liu et al. focused on extracting signatures from networks~\cite{liu2025} that are deeper than the ones achieved by~\cite{carlini2020} and~\cite{canales23}.

%Although most of the literature has focused on developing more powerful model stealing attacks, parallel work has also been conducted on the defense side. Still, it has been observed that most of these defense methods provide only partial protection and often compromise the usability of the model~\cite{jagielski2020high}. 
%The proposed techniques include privacy masking (limiting the sensitivity of output probabilities), output corruption, watermarking models to detect unauthorized use, and information distillation to reduce information leakage \cite{yang2019effectiveness}. However, it has been observed that most of these defense methods provide only partial protection and often compromise the usability of the model \cite{jagielski2020high}. 
%A countermeasure directly targeting the mathematical principles of the cryptanalytic attacks was introduced by Kurian and Aysu~\cite{kurian2025train}. This defense proposes an inference-sensitive training method with a regularization term that minimizes neuronal weight distances to eliminate the uniqueness of neurons. 
%On the other hand, to protect against side-channel model extraction attacks, there are already several proposals, mostly exploring masking countermeasures, see, e.g.,~\cite{9300276,10.1145/3400302.3415649,Dubey_Ahmad_Pasha_Cammarota_Aysu_2021}.

Our work takes a complementary view of this cryptanalytic extraction line. Rather than restricting to a single output component around a critical point, we consider the full vector-valued affine behavior of adjacent linear cells. This reveals a rank-one Jacobian difference whose rows recover the usual neuron signature while its columns expose additional information about later layers. 
A second motivation is numerical: existing extraction techniques rely on highly accurate signature recovery, and finite-precision evaluation can become a significant source of error, particularly in reduced-precision settings such as float32 and float16. The full Jacobian difference provides several redundant observations of the same row direction, suggesting that rank-one approximation or averaging may improve the stability of signature estimation. We investigate both this potential numerical gain and the additional column-side information, while also identifying the limitations of this approach.

\section{Two-Side Signature Recovery}
\label{sec:twoside}

An important property of DNNs with $\Relu$, or another piecewise-linear
activation function, is that the function they realize is piecewise affine
(often called piecewise linear in this context).
We will see how this property allows us to extract the neuron signatures. 

\begin{definition}[Piecewise linear function]
A function $F: \R^n \rightarrow \R^m$ is said to be \textbf{piecewise linear} if there exists a finite collection $\calC =\{C_1, \ldots, C_t \}$ of polyhedra, each described by a finite system of affine inequalities, such that $\bigcup_{i=1}^t C_i = \R^n$ and
$$
    F_{|C_i}(\bm x) = \bm J_{|C_i} \bm x + \bm b_{|C_i},
$$ 
where $F_{|C_i}$ denotes the restriction of $F$ to the set $C_i$ and $\bm J_{|C_i} \in \R^{m \times n}, \bm b_{|C_i} \in \R^{m}.$
In other words, the restriction of $F$ to the sets $C_i$ is an affine function.
The sets $C_i$ are also known as \textbf{linear cells}.
We define the \textbf{rank} of $F_{|C_i}$ as $\Rank(\bm J_{|C_i})$.
If $F$ is continuous and two full-dimensional cells $C_i,C_j$ share a
codimension-one boundary, their affine restrictions agree on that boundary.
Consequently, every point $\bm x$ on the shared boundary satisfies
$$
    (\bm J_{|C_i} - \bm J_{|C_j}) \bm x + \bm b_{|C_i} - \bm b_{|C_j} = 0,
$$
For a ReLU network, such boundary pieces arise when a neuron preactivation is
zero. In the network input space, they are preimages of the critical
hyperplanes introduced in Definition~\ref{def:critical_point} under the
subnetwork preceding the corresponding neuron.
We refer to these preimages as the \textbf{cell boundary}.

\end{definition}

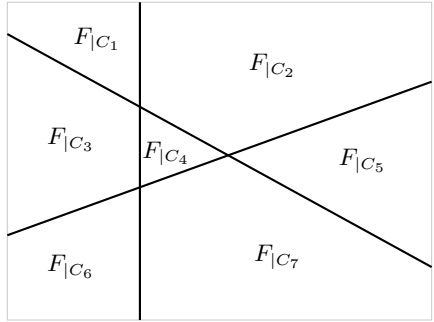
\begin{figure}[htbp]
\centering
\begin{tikzpicture}[scale=0.7]

% outer box just to frame the picture
\draw[thin, gray!40] (-4,-3) rectangle (4,3);

% hyperplanes
\draw[thick] (-4,-1.4) -- (4,1.5);
\draw[thick] (-1.5,-3) -- (-1.5,3);
\draw[thick] (-4,2.4) -- (4,-2);
%\draw[thick] (1.8,-3) -- (1.8,3);

% labels in some cells
\node at (-2.3,2.3) {$F_{\mid C_1}$};
\node at (1, 1.8) {$F_{\mid C_2}$};
%\node at (2.8,2.3) {$F_{\mid C_3}$};

\node at (-2.8,0.4) {$F_{\mid C_3}$};
\node at (-1.0,0.15) {$F_{\mid C_4}$};
\node at (2.7,0) {$F_{\mid C_5}$};

\node at (-2.8,-2) {$F_{\mid C_6}$};
\node at (1.1,-1.8) {$F_{\mid C_7}$};
%\node at (2.7,-2.0) {$F_{\mid C_9}$};

\end{tikzpicture}
\caption{A subdivision of $\mathbb{R}^2$ into linear cells. On each cell $C_i$, the restriction $F_{\mid C_i}$ of $F$ is affine.}
\label{fig:linear-cells}
\end{figure}

The composition of affine functions and piecewise linear functions results in piecewise linear functions whose linear cells are delimited by portions of hyperplanes.

Recall from Definition \ref{def:DNN} that the function implemented by a neural network can be described as
\begin{equation}
\tag{\ref{eq:DNN}}
F(\bm x) = A^{(\ell)} \circ \bm \sigma \circ \cdots \circ A^{(2)} \circ \bm \sigma \circ A^{(1)}(\bm x),
\end{equation}
where $A^{(k)}(\bm x) = \bm W^{(k)} \bm x  + \bm b^{(k)}$ for some weight matrices $\bm W^{(k)} \in \R^{n_k \times n_{k-1}}$ and bias vectors $\bm b^{(k)} \in \R^{n_k}$, hence they are affine functions.

Let $\bm \sigma$ be the component-wise application of $\Relu$, which is a piecewise linear function; then the function $F$ is also a piecewise linear function. 
The borders of the linear cells of $F$ are determined by the activation function.

Once we fix a particular input, this will determine a particular configuration of active-inactive neurons.
Each of these patterns can be represented by substituting the function $\bm \sigma$ at layer $k$ with a diagonal matrix whose diagonal entries are $1$ if $\bm w_{i,*}^{(k)\intercal} \bm h^{(k-1)}(\bm x) + b_i^{(k)} > 0$ (i.e., the corresponding neuron is active) and $0$ otherwise.
For a linear cell $C$, we denote this diagonal matrix by $\bm D_C^{(k)}\in\R^{n_k\times n_k}$.
Each point in the input space $\bm x \in \R^{n_0}$ lies within a linear cell or on the border between two or more cells; inside the same linear cell, all the diagonal matrices $\bm D_C^{(k)}$ are constant.
As the borders are constituted by portions of hyperplanes having measure zero, a random point $\bm x$ will be inside a linear cell with probability $1$.\footnote{In practice, this probability will be only close to $1$ as we cannot have infinite precision.}

We denote the linear cell containing the vector $\bm x_0$ by $ C_0$, and the restriction of the function $F$ to the cell $C_0$ by $ F_{| C_0}$. 
Locally (if we restrict the domain to  $ C_0 \subseteq \R^{n_0}$), the function $F$ is a composition of affine functions, and it is therefore affine.
Using this notation, we define:
\begin{align}
\bm J_{|C_0} &= \bm W^{(\ell)} \bm D_{C_0}^{(\ell-1)} \cdots \bm D_{C_0}^{(1)} \bm W^{(1)} \in \R^{n_\ell \times n_0} \\
\bm L_{i,C_0} &= \bm W^{(\ell)}\bm D_{C_0}^{(\ell-1)} \cdots \bm W^{(i+1)}, i = 1,\dots, \ell-1\\
\bm b_{|C_0} &= \sum_{i=1}^{\ell-1} \bm L_{i,C_0} \bm D_{C_0}^{(i)}  \bm b^{(i)} + \bm b^{(\ell)}
\end{align}
the restricted function $F_{|C_0}$ can be expressed as
\begin{equation}
\label{eq:F=J+b}
    F_{| C_0} (\bm x) = \bm J_{|C_0} \bm x + \bm b_{|C_0}.
\end{equation}
Informally, we will refer to the rank of the matrix $\bm J_{|C_0}$ as the rank of $F_{|C_0}.$
The affine function $F_{|C_0}: \R^{n_0} \rightarrow \R^{n_\ell}$, can be reconstructed choosing $n_0+1$ affine independent points in $C_0$ where $n_0$ is the dimension of the input space.
Provided we have access to the evaluation of the function at these points, and that all these points belong to the same linear cell $C_0$, it is always possible to determine the function $F_{|C_0}$ and the matrix $\bm J_{|C_0}.$
Another interpretation of the matrix $\bm J_{|C_0}$ is that it is the Jacobian of the function $F_{|C_0}$ that is
\begin{equation}
\label{eq:G_Jacobian}
    \bm J_{|C_0} = 
    \begin{pmatrix}
\frac{\partial F_{1|C_0,}}{\partial \bm e_1} & \frac{\partial F_{1|C_0}}{\partial \bm e_2} & \cdots & \frac{\partial F_{1|C_0}}{\partial \bm e_{n_0}} \\
\frac{\partial F_{2|C_0,}}{\partial \bm e_1} & \frac{\partial F_{2|C_0}}{\partial \bm e_2} & \cdots & \frac{\partial F_{2|C_0}}{\partial \bm e_{n_0}} \\
\vdots & \vdots & \ddots & \vdots \\
\frac{\partial F_{n_{\ell}|C_0,}}{\partial \bm e_1} & \frac{\partial F_{n_{\ell}|C_0}}{\partial \bm e_2} & \cdots & \frac{\partial F_{n_{\ell}|C_0}}{\partial \bm e_{n_0}} 
\end{pmatrix},
\end{equation}
where $\frac{\partial F_{i|C_0,}}{\partial \bm e_j}$ is the partial derivative in direction $\bm e_j$ of the $i$-th component of the function $F_{|C_0}$ and we denote by $\bm e_j$ the unit vector of the standard basis having $j$-th component $1$ and being zero elsewhere.

This simple fact becomes crucial when we compare the function in two adjacent linear cells, that is, two linear cells separated by a portion of a cell boundary.
Let $C_0$ and $C_1$ be two adjacent linear cells that differ only in the activation of the $j$-th neuron of the $i$-th layer.
The diagonal matrices satisfy $\bm D_{C_0}^{(i)} - \bm D_{C_1}^{(i)} = \pm \bm E_{j,j}$ where $\bm E_{j,j} \in \R^{n_i \times n_i}$ is the matrix that is $0$ in each entry except in the position $(j,j)$ where it is $1.$

It is useful to introduce some notation for the matrix products before and after the activation at layer $i$.
For a linear cell $C_0$, define
\begin{align*}
    \bm R_{i,C_0} &= \bm W^{(i)} \bm D_{C_0}^{(i-1)} \bm W^{(i-1)} \cdots \bm D_{C_0}^{(1)} \bm W^{(1)} \in \R^{n_i\times n_0}, \\
    \bm L_{i,C_0} &= \bm W^{(\ell)} \bm D_{C_0}^{(\ell-1)} \bm W^{(\ell-1)} \cdots \bm W^{(i+1)} \in \R^{n_\ell\times n_i}.
\end{align*}

We can use this to factor the matrix $\bm J_{|C_0}$ as  
%\[
%\bm J_{|C_0} = 
%\underbrace{\bm W^{(\ell)} \bm \sigma_{|C_0}^{(\ell-1)} \cdots \bm \sigma_{| \bm x_0}^{(i+1)} \bm W^{(i+1)}}_{\bm T_{|C_0}^{(\ell - i)}}
%\bm \sigma_{|C_0}^{(i)}
%\underbrace{\bm W^{(i)}_{|C_0} \bm \sigma_{|C_0}^{(i-1)} \cdots \bm \sigma_{|C_0}^{(1)} \bm W^{(1)}}_{\bm H^{(i)}_{|C_0}}
%\]
\begin{equation}
\label{eq:J=TsH}
    \bm J_{|C_0} =  \bm L_{i,C_0} \bm D_{C_0}^{(i)}\bm R_{i,C_0}.
\end{equation}

The only difference between $\bm J_{|C_0}$ and $\bm J_{|C_1}$  is in a single activation of $i$-th layer. Without loss of generality, we label the cells so that this neuron is active in $C_0$ and inactive in $C_1$. Therefore, $\bm D_{C_0}^{(i)}-\bm D_{C_1}^{(i)}=\bm E_{j,j}$,
and the difference between the two Jacobians is given by
\begin{equation}
\label{eq:Signature_as_rank1}
    \bm \Delta(C_0, C_1) = \bm J_{|C_0} - \bm J_{| C_1} = \bm L_{i,C_0} \bm E_{j,j} \bm R_{i,C_0} = \bm c_{*,j} \bm r_{j,*}^{\intercal} \in \mathbb{R}^{n_{\ell} \times n_0},
\end{equation}
where $\bm c_{*,j}$ is the $j$-th column of $\bm L_{i,C_0}$ and $\bm r_{j,*}^\intercal$ is the $j$-th row of $\bm R_{i,C_0}$.

Each row of the matrix $\bm \Delta(C_0, C_1)$ is a scalar multiple of the row vector $\bm r_{j,*}^\intercal$, while each column is a scalar multiple of the column vector $\bm c_{*,j}$. 
In terms of Definition \ref{def:signature}, every row of $\bm \Delta(C_0, C_1)$ has the same row-signature and every column has the same column-signature. 
There are two special cases in which we can exploit this decomposition to recover the row-signatures or the column-signatures of the matrices $\bm W^{(i)}$.
\begin{enumerate}
\item In the first case, the activation patterns for $C_0$ and $C_1$ differ for the $j$-th neuron in the first layer.  Since $\bm R_{1,C_0} = \bm W^{(1)},$  each row of the matrix $\bm \Delta(C_0, C_1)$ is a scalar multiple of the $j$-th row of $\bm W^{(1)}$.
This is exactly the same case exploited in \cite{carlini2020} and its follow-up works.
The connection is even more evident if we recall that $\bm J_{|C_0}$ is the Jacobian of the local affine function $F_{|C_0}(\bm x).$ 

\item In the second case (new), the activation patterns differ in the $j$-th neuron of the last hidden layer. Since $\bm L_{\ell-1,C_0} = \bm W^{(\ell)},$ the columns of $\bm \Delta(C_0, C_1)$ are scalar multiples of the $j$-th column of $\bm W^{(\ell)}$.
\end{enumerate}
As already stated, the first case was exploited to collect all the signatures of the first layer and from there attack the second layer, and so on.
Then, a natural question arises: Can we follow a similar approach, collecting the column signature of the last layer, and proceed backward towards the first?
If successful, then the two approaches could be easily combined, reconstructing the function from both ends.

\subsection{Finding adjacent linear cells}
\label{sec:find_adjacent}

To determine signatures, it is essential to find adjacent linear cells. 
Consider an input $\bm x_0$ and a unit vector $\bm v$, we can choose a step length $\delta > 0$ and consider the sequence of points $S(\bm x_0, \bm v) = \{ \bm x_t = \bm x_0 + t \delta \bm v \}_{t \in \mathbb Z}$.
We denote the finite differences of consecutive elements of this sequence as
\begin{equation} \label{eq:partial_d}
    D_{\bm v,\delta}F(\bm x_t) = 
    \frac{F(\bm x_{t+1}) - F(\bm x_{t})}{\delta} = \frac{F(\bm x_0 + (t+1)\delta\bm v) - F(\bm x_{0} + t\delta \bm v)}{\delta}.
\end{equation}
If the interval $[\bm x_t,\bm x_{t+1}]$ is contained in a single linear cell $C_0$ of the piecewise linear function $F$, then $D_{\bm v,\delta}F(\bm x_t)$ coincides with the directional derivative of $F$ along $\bm v$ at $\bm x_t$, which we denote by $D_{\bm v}F(\bm x_t)$. In particular,
\[
    D_{\bm v,\delta}F(\bm x_t)
    =
    D_{\bm v}F(\bm x_t)
    =
    \bm J_{|C_0}\bm v.
\]
If the interval crosses a cell boundary, $D_{\bm v,\delta}F(\bm x_t)$ remains a finite difference and does not need to coincide with the directional derivative at $\bm x_t$.

Within the linear cell $C_0$ the function $F$ can be expressed as $F_{|C_0}(\bm x) = \bm J_{|C_0} \bm x + \bm b_{|C_0}$, its finite difference along $\bm v$ is the constant vector $D_{\bm v,\delta}F(\bm x) = \bm J_{|C_0} \bm v$. 
Suppose $\bm x_0 \in C_0$, and let $\bm x_{t_1 -1} \in C_0$ be the last element in the sequence $S(\bm x_0, \bm v)$ that belongs to the same cell, we will have 
$$
D_{\bm v,\delta}F(\bm x_0) = D_{\bm v,\delta}F(\bm x_1) = \cdots = D_{\bm v,\delta}F(\bm x_{t_1-2}) = \bm J_{|C_0} \bm v.
$$ 

As $\bm x_{t_1} \in C_1$ while $\bm x_{t_1 - 1} \in C_0$ there will be some value of $0 \leq \lambda <1$ for which $\bm x_{t_1 -1} + \lambda \delta \bm v$ lies exactly on the cell boundary dividing $C_0$ from $C_1$,
then 
\begin{equation}
\label{eq:part_xt1}
    D_{\bm v,\delta}F(\bm x_{t_1-1}) = (\lambda \bm J_{|C_0} + (1-\lambda)\bm J_{|C_1}) \bm v \neq  \bm J_{|C_0} \bm v,
\end{equation}
except in the unlikely case where $\bm J_{|C_0} \bm v= \bm J_{|C_1} \bm v.$

From $\bm x_{t_1}$ onward, the finite difference $D_{\bm v,\delta}F(\bm x_{t_1})$ will be again a constant vector given by $\bm J_{|C_1} \bm v$ as long as we stay within $C_1$.
If the sequence leaves $C_1$ at a certain point $x_{t_2}$, we will detect this transition by observing another discontinuity in the value of $D_{\bm v,\delta}F$.

\iffalse
A method to do so is to select two inputs $\bm x_0, \bm x_1$ and move along a line that connects them, evaluating the function $F$ at regular intervals. \todo[color=cyan]{Maybe would be simpler to present as: choose an input  $\bm x$ and a direction given by a vector $\bm v$ of norm $||v||=1$.?}
The line can be parametrized by $0\leq\lambda \leq 1$
as 
$\lambda \bm x_0  + (1-\lambda) \bm  x_1.$ 
More formally, define $\bm v =\bm  x_1 - \bm x_0$ and consider the points 
$\bm x_0 + \frac{t}{n} \bm v$ for $t \in [0,\ldots,n]$.
If two consecutive points are in the same linear cell $C_0$, then we have:
\begin{equation} \label{eq:partial_d}
    \frac{F(\bm x_0 + \frac{t+1}{n}\bm v) - F(\bm x_0 + \frac{t}{n}\bm v)}{\frac{1}{n} ||\bm v||} = \frac{\partial F}{\partial \bm v}\left (\bm x_0 + \frac{t}{n}\bm v \right ).
\end{equation}
As the derivative of an affine function is constant, this vector will be constant inside each linear cell.
In particular, we have
\begin{equation} 
    \frac{F(\bm x_0 + \frac{t+1}{n}\bm v) - F(\bm x_0 + \frac{t}{n}\bm v)}{\frac{1}{n} ||\bm v||} = 
    \bm G_{|C_0} \frac{\bm v}{||\bm v ||}.
\end{equation}
However, if $\bm x_1 \in C_1 \neq C_0$, then
moving along the line, at some point we will leave $C_0$ and enter the adjacent linear cell $C_1$.
In this case, we will, with high probability, notice a change in the derivative.
\fi

Denote by $\bm d$ the difference between the directional derivatives of the affine restrictions of $F$ to the two cells:
\[
\bm d = \bm J_{|C_0} \bm v - \bm J_{|C_1} \bm v = \bm \Delta(C_0, C_1) \bm v.
\]
With a dichotomy, it is possible to identify the location of the critical point where this discontinuity is located \cite{carlini2020}.

Alternatively \cite{canales23}, this point can be located as $\bm x_{t_1 - 1} + \lambda \delta \bm v$, solving for lambda in equation (\ref{eq:part_xt1}) noticing that the values of $\bm J_{|C_0} \bm v$ and $\bm J_{|C_1} \bm v$ are known as they can be calculated from the evaluations of $F$.

Observe that \cite{carlini2020} and subsequent works restrict the function $F$ to a single component.
In that case, the finite difference is a scalar, corresponding to the same component of the vector $D_{\bm v,\delta}F(\bm x_t)$.
Transitions from one linear cell to another are detected when this scalar changes.
Clearly, it is possible for two points $\bm x_{t_0}$ and $\bm x_{t_1}$ to be in different linear cells, where the vectors $D_{\bm v,\delta}F\left (\bm x_{t_0} \right ) \neq D_{\bm v,\delta}F\left (\bm x_{t_1} \right )$, while the same vectors are equal in the component to which $F$ has been restricted.
In this case, the method in \cite{carlini2020} fails to distinguish $C_0,C_1$ as two distinct linear cells.
This event has measure zero, but it might happen that the two scalars are so close that they cannot be reliably detected.

\subsection{Extracting a row signature}
In \cite{carlini2020} and following works, signatures are extracted by comparing the derivative on both sides of a cell boundary on a single component $q$ of the output vector $F_q(\bm x) = [F(\bm x)]_q \in \R$.
Before describing our technique, we will briefly describe how signatures are recovered in \cite{canales23}.
Let $\bm e_i$ be the standard basis unit vector that has $1$ in its $i$-th entry and zero elsewhere, and let $\bm x^*$ be a point on the cell boundary.
They define $ \alpha_{i,+} = D_{\bm e_i,\delta}F_q\left (\bm x^* + \delta \bm e_i\right ) $ and $ \alpha_{i,-} = D_{\bm e_i,\delta}F_q\left (\bm x^* - \delta \bm e_i\right )$.
For the first layer, they consider the signature
$$
\left( \frac{\alpha_{1,+} - \alpha_{1,-}}{\alpha_{r,+} - \alpha_{r,-}}, \frac{\alpha_{2,+} - \alpha_{2,-}}{\alpha_{r,+} - \alpha_{r,-}}, \ldots, \frac{\alpha_{n_0,+} - \alpha_{n_0,-}}{\alpha_{r,+} - \alpha_{r,-}}
\right ),
$$ 
and by choosing $r = 1$, they will obtain a signature whose first component is $1$.
It can be verified that, if no other neuron flips on the points used to estimate the values of $\alpha_{i,+},\alpha_{i,-}$, the signature obtained is a scalar multiple of the weight of the neuron.

Once the weights of a neuron $\eta_i^{(k)}$ are recovered,  its bias can be determined from Equation (\ref{eq:critical_point}) in Definition \ref{def:critical_point}.
For convenience, we recall that
\begin{equation}\label{eqqq}
    - \bm w_{i,*}^{(k) \intercal} h^{(k-1)}(\bm x)  = b_i^{(k)},
\end{equation}
where $h^{(k-1)}(\bm x)$ is a point that belongs to the critical hyperplane associated with the neuron $\eta_i^{(k)}$ and $\bm x$ is a point that belongs to a cell boundary of two adjacent cells differing by the activation of $\eta_i^{(k)}$.
Further details on the recovery of such a point, and hence of the bias, are given in Section \ref{sec:bias}.   

\paragraph{Our method: extraction from the Jacobian}
In the discussion following (\ref{eq:Signature_as_rank1}), we already observed how the Jacobians of two adjacent linear cells $C_0,C_1$ differ by a rank $1$ matrix.
If the neuron that flipped between $C_0$ and $C_1$ is in the first layer, the rows of $\bm \Delta(C_0, C_1) = \bm J_{|C_0} - \bm J_{|C_1}$ will all be scalar multiples of the row in the matrix $\bm W^{(1)}$ corresponding to the weights of the neuron that flipped.

To compute the Jacobians $\bm J_{|C_i}$ we need to find a point $\bm x^* \in C_i$ that is at a distance of at least $\delta$ from any cell boundaries.
In this way, we can query the model on $\bm x^*$ and on neighbor points $\bm x^* + \delta \bm v_j$, where $\bm v_1, \ldots \bm v_{n_0}$ are linearly independent unit vectors, granted that all these points are still in the same linear cell.  

Assume that both $\bm x^* , \bm x^* + \delta \bm v_j \in C_i $ and compute
$\bm y_j = F(\bm x^* + \delta \bm v_j) - F(\bm x^*) = \delta \bm J_{|C_i} \bm v_j,$ we call $\bm Y$ the matrix whose columns are given by the vectors $\bm y_j$ and $\bm V$ the matrix whose columns are the vectors $\bm v_j$. 
Then
$$
    \bm J_{|C_i} = \frac{1}{\delta} \bm Y \bm V^{-1},
$$
a natural choice is to choose $\bm v_j = \bm e_j$ the standard basis, in this way the matrix $\bm V$ is the identity.

It is crucial that all the points $\bm x^* + \delta \bm v_j$ belong to the same cell $C_i$.
In Section \ref{sec:find_adjacent} we introduced the sequence $S(\bm x_0, \bm u)$.
Let $\bm x_0 \in C_0$, we name $\bm x_{t_1}$ the first element in $C_1$, $\bm x_{t_2}$ the first element in $C_2$ and so on.
A first good candidate is given by $\bm x^*= \frac{\bm x_{t_1} + \bm x_{t_{2} - 1}}{2},$ more generally, a good point to compute $\bm J_{|C_i}$ is given by the midpoint between $\bm x_{t_i}$ and $\bm x_{t_{i+1} - 1}$.
This point will be almost as far as possible from the boundaries of cell $C_1$ along the line defined by $\bm x_0$ and the vector $\bm u$.
Unfortunately, this will not prevent the same point from being close to any other cell boundary.
A practical consistency test for $\bm x^*$ and
$\bm x^*+\delta\bm v_j$ is to compare finite differences along
one or more probe directions $\bm u_s$:
\[
\left\|
\frac{F(\bm x^*+\mu\bm u_s)-F(\bm x^*)}{\mu}
-
\frac{F(\bm x^*+\delta\bm v_j+\mu\bm u_s)
      -F(\bm x^*+\delta\bm v_j)}{\mu}
\right\|\leq\tau .
\]
In exact arithmetic $\tau=0$; using several probe directions reduces the chance of accepting points from different cells.

\paragraph{Sign}
Similar to other techniques, our signatures will miss a global sign,
in~\cite{carlini2020},~\cite{canales23}, and~\cite{liu2025}, several techniques are proposed to recover this last missing information about each neuron.
These techniques can be applied to our case in the same way as reconstructing the first layer.

\subsection{Attacking deeper layers}
Once the first layer is reconstructed, we can proceed to attack the second.
The process will be the same for each subsequent layer, so we will assume that we have successfully reconstructed the first $k-1$ layers and that we want to attack the $k$-th layer.

In our model, a signature is recovered as normalized rows and columns of $\bm \Delta(C_0, C_1) = \bm J_{|C_0} - \bm J_{|C_1}$, where $C_0, C_1$ are adjacent linear cells (i.e., cells differing only by one activation).
Suppose the neuron whose activation determines the separation between $C_0, C_1$ is the $j$-th neuron of the $k$-th layer and that we have already successfully reconstructed the first $k-1$ layers. From (\ref{eq:Signature_as_rank1}) we have
$$
    \bm \Delta(C_0, C_1) = \bm J_{|C_0} - \bm J_{|C_1} = 
    \bm L_{k,C_0} \bm E_{j,j} \bm R_{k,C_0} = \bm L_{k,C_0} \bm E_{j,j} \bm W^{(k)}  \bm D_{C_0}^{(k-1)} \bm R_{k-1,C_0},
$$
Notice that, the matrix $\bm L_{k,C_0} \bm E_{j,j} \bm W^{(k)}$ will be a rank $1$ matrix whose rows are all scalar multiples of $\bm w_{j,*}^{(k)\intercal}$, the $j$-th row of $\bm W^{(k)}$ and whose columns are all scalar multiples of the $j$-th column of $\bm L_{k,C_0}$.

The matrix $\bm R_{k-1,C_0}$ is known, as we have already reconstructed the first $k-1$ layers.
Then, if this matrix has rank $n_{k-1}$, it is possible to find an  invertible sub-matrix $n_{k-1}\times n_{k-1}$ and we can recover the matrix $\bm L_{k,C_0} \bm E_{j,j} \bm W^{(k)} \bm D_{C_0}^{(k-1)}$.
Notice that this last operation hides two problems.
One is that the matrix $\bm R_{k-1,C_0}$ might not have the desired rank, and the other is that the columns of $\bm L_{k,C_0} \bm E_{j,j} \bm W^{(k)} \bm D_{C_0}^{(k-1)}$ will be zero in correspondence with the inactive neurons of the $(k-1)$-th layer.
The first problem is known as insufficient rank signatures and the second as partial signatures.
We will treat them separately in the following two paragraphs.

\paragraph{Partial signatures}
Partial signatures emerged in a slightly different way in \cite{carlini2020} and subsequent works.
The solution proposed in \cite{carlini2020} can be used for partial signatures extracted from the Jacobian.
For the sake of clarity, we will briefly describe the method proposed in \cite{carlini2020} to merge partial signatures in our framework.

Under the assumption that $\bm R_{k-1,C_0} \in \R^{n_{k-1} \times n_0}$ has rank $n_{k-1}$, the signature we are recovering will be the $j$-th row of $\bm W^{(k)}$, denoted by $\bm w_{j,*}^{(k)\intercal}$, multiplied by the matrix $\bm D_{C_0}^{(k-1)}$.
Unless all the neurons in the $(k-1)$-th layer are active, the row vector $\bm w_{j,*}^{(k)\intercal} \bm D_{C_0}^{(k-1)}$ will have some zeros corresponding to the zeros in $\bm D_{C_0}^{(k-1)}$.
Notice that, if we observe a partial signature relative to the same $j$-th neuron of the $k$-th layer from a different pair of adjacent cells $C_i, C_{i+1},$ the matrix $\bm D_{C_i}^{(k-1)}$ will likely reveal some coordinates that were previously zero and mask others with zeros.
In \cite{carlini2020}, a technique was already proposed to identify and merge partial signatures coming from the same neuron.
This method can be illustrated with a simple example.
Consider two partial signatures $\bm v = (v_1, v_2, 0, v_4, 0)^\intercal$ and $\bm u = (u_1, u_2, u_3, 0, u_5)^\intercal$ that have two non-zero entries in the first two components.
If the two vectors are a scaled and masked version of the same row vector $\bm w^{(k)}_{j,*}$, then $\frac{u_1}{u_2} = \frac{w_{j,1}^{(k)}}{w_{j,2}^{(k)}} = \frac{v_1}{v_2}$.
This simple test gives a criterion to establish whether two signatures are relative to the same neuron and can be merged.
The implicit assumption is that different neurons will have different ratios between the overlapping components. 
To merge the two signatures, each vector is normalized by one of their common nonzero components, and the missing entries of one signature are filled using the corresponding nonzero entries of the other. 
In the example above, normalizing with respect to the first component yields
$$
\left (\frac{v_1}{v_1},\frac{v_2}{v_1},\frac{u_3}{u_1}, \frac{v_4}{v_1},\frac{u_5}{u_1} \right ).
$$

\paragraph{Insufficient rank signatures}

The problem of insufficient rank signatures arises when the matrix $\bm R_{k-1,C_0}$ has rank lower than $n_{k-1}$.
This problem is more severe than partial signatures as it can prevent the recovery of the signature of a neuron in the $k$-th layer.
A solution to this problem was proposed in \cite{liu2025} combining linear equations coming from distinct portions (distinct pairs of adjacent cells) of cell boundaries relative to pairs of adjacent cells differing in the activation of the same neuron.
Instead of explaining the method in \cite{liu2025}, we will describe a similar method adapted to our framework.

Let us denote by $\bm \Delta_1, \bm \Delta_2, \ldots, \bm \Delta_m$ the matrices $\bm \Delta_i = \bm \Delta(C_{i,0}, C_{i,1}) = \bm J_{|C_{i,0}} - \bm J_{|C_{i,1}}$ relative to $m$ distinct pairs of adjacent cells $C_{i,0}, C_{i,1}$ that differ only in the activation of the same neuron in the $k$-th layer.
All these matrices will have rank $1$, and each will satisfy an equation of the form
$$
\bm \Delta_i = \bm v_{|C_{i,0}} \bm w^{(k) \intercal}_{j,*} \bm M_{|C_{i,0}},
$$
where $\bm v_{|C_{i,0}}$ is the $j$-th column of $\bm L_{k, C_{i,0}} $ and  $\bm M_{|C_{i,0}} = \bm D_{C_{i,0}}^{(k-1)} \bm R_{k-1, C_{i,0}} \in \R^{n_{k-1} \times n_{0}}$.
Noticing that both $\bm \Delta_i$ and $\bm v_{|C_{i,0}} \bm w^{(k) \intercal}_{j,*}$ are rank $1$ matrices, we can write the equation above as a system of linear equations in the unknown $\bm w^{(k) \intercal}_{j,*}$, ignoring scalar multipliers as
$$
\bm M_{|C_{i,0}}^\intercal \bm w^{(k)}_{j,*} = \lambda \bm d_i
$$
where $\bm d_i$ is a row of the matrix $\bm \Delta_i$ up to scalar multiplication.
Treating $\bm w^{(k)}_{j,*}$ as the unknown, the solution to this system will be given by the vector space $\mathcal{V}_i = \langle \bm w_i \rangle + \ker(\bm M_{|C_{i,0}}^\intercal)$, where $\bm w_i$ is a particular solution to the system.
As $\bm w^{(k)}_{j,*}$ is a solution to all the systems, it will belong to the intersection of all the vector spaces $\mathcal{V}_i$.
Notice that the matrix $\bm M_{|C_{i,0}}^\intercal$ has zero columns corresponding to the inactive neurons of the $(k-1)$-th layer, so the kernel of $\bm M_{|C_{i,0}}^\intercal$ contains at least the space spanned by $\bm e_j$ for each inactive neuron $j$ of the $(k-1)$-th layer.
We can recover a partial signature if the kernel of the matrix $\bm M_{|C_{i,0}}^\intercal$ from which we remove the zero columns is trivial.
If this kernel is not trivial, we can intersect with a different solution, keeping in mind to remove the zero columns where the two matrices share zero columns and keep the zero columns where only one of the two matrices has a zero column.

A priori, we do not know if $\bm \Delta_i$ and $\bm \Delta_j$ are relative to the same neuron, but we can test them in the same way proposed in \cite{liu2025}.
In particular, the intersection of the two vector spaces $\mathcal{V}_i \cap \mathcal{V}_j$ is expected to be trivial if the sum of their dimension is smaller than the number of variables (i.e., is smaller than the number of surviving columns in the two matrices).
If the intersection has dimension $1$, most likely we have recovered the signature even if the matrices $\bm M_{|C_{i,0}}$ and $\bm M_{|C_{j,0}}$ were rank deficient.

\subsection{Extracting bias}\label{sec:bias}
Similar to what is done in \cite{carlini2020}, after we successfully recovered a complete signature for a neuron $\eta_i^{(k)}$, it is possible to recover its bias from the following equation (deduced from Relation (\ref{eqqq})):
\begin{equation}
\label{eq:bias_extraction}
    - \bm w_{i,*}^{(k) \intercal} h^{( k-1)}( \bm x^* ) = b_i^{(k)}
\end{equation}
where $h^{(k-1)}( \bm x^* ) \in \R^{n_{k - 1}}$ is a point that belongs to the critical hyperplane of $\eta_i^{(k)}.$
Since a signature determines the weight row only up to scale and sign, we normalize the recovered row as
$$
\widehat{\bm w}_{i,*}^{(k)}
=
\frac{\bm w_{i,*}^{(k)}}{\lVert\bm w_{i,*}^{(k)}\rVert_2}.
$$
Consequently, Equation~\eqref{eq:bias_extraction} recovers the normalized bias
$$
\widehat b_i^{(k)}
=
\frac{b_i^{(k)}}{\lVert\bm w_{i,*}^{(k)}\rVert_2},
$$
up to the selected sign, rather than the physical parameter \(b_i^{(k)}\).

Suppose we have already reconstructed the neural network up to layer $k$ and let $\bm x_{1}, \bm x_{2}$ be two points belonging to two adjacent cells differing in the activation of neuron $\eta_i^{(k)}$.
That is, the two cells are divided by the portion cell boundary $(h^{(k-1)})^{-1}(\calH_{i}^{(k)})$, where $\calH_{i}^{(k)} \subseteq \R^{n_{k - 1}}$ is the critical hyperplane relative to the neuron $\eta_i^{(k)}$ (see Definition \ref{def:critical_point}).
We assume that the segment $[\bm x_1,\bm x_2]$ crosses exactly one cell boundary, at a point $\bm x^* \in (h^{(k-1)})^{-1}(\calH_{i}^{(k)})$.

Since $\bm x_1$ and $\bm x_2$ lie on opposite sides of the same critical boundary, one can search along the segment joining them for the point at which the local affine behavior of $F$ changes.
The continuity of $F$ guarantees that the two affine pieces agree at the critical point.
Let
\begin{equation}
\bm x^* = \bm x_1 + t^*(\bm x_2-\bm x_1),
\qquad t^* \in [0,1],
\end{equation}
and denote by $\bm J_1$ and $\bm J_2$ the Jacobians of $F$ in the cells containing $\bm x_1$ and $\bm x_2$, respectively. The continuity of $F$ ensures that
\[
F(\bm x_1)
+ t^* \bm J_1(\bm x_2-\bm x_1)
=
F(\bm x_2)
- (1-t^*)\bm J_2(\bm x_2-\bm x_1).
\]
Rearranging, we get
\[
t^* (\bm J_1 - \bm J_2)(\bm x_2-\bm x_1)
=
F(\bm x_2) - F(\bm x_1)
- \bm J_2(\bm x_2-\bm x_1).
\]
Denoting by $\bm y$ the vector on the right-hand side of the equation and by $\bm a$ the vector $(\bm J_1 - \bm J_2)(\bm x_2-\bm x_1),$ we get the simpler equation $t^* \bm a = \bm y$.

In \cite{carlini2020}, a single output coordinate (for example the first), is used to determine $t^*$ as $t^* = \frac{y_1}{a_1}.$ This formula is correct if all computations are exact. Since we typically work with
finite numerical precision, the equation $t^* = \frac{y_1}{a_1}$ might have no exact solution. In this case,
it is usually better to determine the least-squares solution
$$
\widehat{t}
=
\arg\min_{t}
\|t\bm a-\bm y\|_2^2,
$$
which is given by
$$
\widehat{t}
=
\frac{\bm a^\top \bm y}{\bm a^\top \bm a}.
$$

Hence, coordinates with larger \(|a_r|\) contribute more strongly 
while coordinates for which \(a_r\) is close to zero are naturally downweighted, since small numerical perturbations in \(y_r\) are strongly amplified in the ratio \(y_r/a_r\).

\subsection{Numerical precision}

\paragraph{Enhance precision}
Due to numerical precision, the recovery of the signatures will always have some error.
These errors will accumulate when we explore deeper layers; this affects both the fidelity of the reconstructed model and the capacity to reliably extract deep layers.

In previous works, signatures were collected as single row-signatures and grouped by similarity, fixing a threshold distance.
In our case, the matrix 
\[\bm \Delta(C_i,C_j) = \bm J_{|C_i} - \bm J_{|C_j}\] will contain rows (or columns) all referring to the same signature.
Rather than extracting these row-signatures independently, we apply Singular Value Decomposition (SVD) to $\bm \Delta(C_i,C_j)$ and estimate their common direction from its dominant right singular vector. 
This allows the information contained in all output components to be combined into a single signature estimate, improving the precision of our extraction.
If the entries correspond to non-adjacent cells, the matrix is no longer approximately rank-$1$; this is reflected in the presence of multiple significant singular values. 
We propose two further techniques to improve the numerical precision.

\paragraph{Adaptive steps}
Inside a linear cell $C$, the Jacobian $\bm J_{|C}$ is constant, so a directional derivative can be computed using any nonzero interval contained entirely in that cell.
In particular, for a unit vector $\bm v$ and $h_+,h_- \geq 0$ with $h_+ + h_- > 0$, if $[\bm x-h_-\bm v,\bm x+h_+\bm v] \subseteq C$, then in exact arithmetic
\begin{equation}
\label{eq:chord-derivative}
    \frac{F(\bm x+h_+\bm v)-F(\bm x-h_-\bm v)}{h_+ + h_-}
    = \bm J_{|C}\bm v.
\end{equation}
If each endpoint evaluation has an additive error of norm at most $\eta$, its contribution to the derivative error is bounded by $2\eta/(h_+ + h_-)$, before accounting for rounding in the subtraction and division.
This motivates using larger steps.

This principle was also used by Noman et al.~\cite[Sec.~4.1]{noman2026rounding}
in a study of deterministic output rounding as a defense against
cryptanalytic extraction.
Their step-spacing attack adaptively enlarges symmetric finite-difference
intervals, subject to a local linearity test, to recover directional
derivatives despite output rounding.
The defense rounds only the returned outputs, while retaining
full-precision weights and forward computation.
Our construction applies the same large-step principle to
vector-valued Jacobian estimation, allowing asymmetric intervals and
using the baseline Jacobian for consistency checks.

We first obtain a baseline estimate $\widehat{\bm J}_{|C}$ using the small-step procedure described above.
Once this estimate is available, we can test whether a candidate point $\bm z$ is locally consistent with the same Jacobian.
For a unit probe direction $\bm v$ and a small step $\delta$, we check whether
\begin{equation}
\label{eq:chord-consistency}
    \left\|F(\bm z+\delta\bm v)-F(\bm z)
    -\delta\widehat{\bm J}_{|C}\bm v\right\| \leq \tau,
\end{equation}
where $\tau$ accounts for both evaluation errors and uncertainty in the baseline estimate.
A single probe requires at most two additional oracle queries; several probe directions (and correspondingly more queries) can be used to reduce the risk of accepting a point with a different local Jacobian.
This is a heuristic consistency test, not a certificate of cell membership: distinct cells can agree along the tested directions, and endpoint tests alone cannot exclude intervening boundaries.
Its reliability also depends on the quality of the baseline estimate.

Starting from a point $\bm x$ in $C$, we extend a segment along each coordinate direction $\bm e_i$.
On the positive side, we test $\bm x+\delta\bm e_i$, $\bm x+2\delta\bm e_i$, doubling the step at each iteration, stopping at the first failed consistency test or at a prescribed search limit.
We retain the last accepted displacement $\bm x + 2^{k_{i}^+} \delta \bm e_i$ in one direction and repeat the same procedure independently in the opposite direction, retaining $\bm x - 2^{k_{i}^-} \delta \bm e_i$.
The two points need not be equally distant from $\bm x$.
For convenience we rename $h_{i,+} = 2^{k_{i}^+} \delta$ and $h_{i,-} = 2^{k_{i}^-} \delta$, we then estimate the $i$-th column of the Jacobian by
\begin{equation}
\label{eq:chord-jacobian-column}
    \widehat{\bm J}_{|C}\bm e_i
    = \frac{F(\bm x+h_{i,+}\bm e_i)-F(\bm x-h_{i,-}\bm e_i)}
    {h_{i,+}+h_{i,-}},
\end{equation}
provided the total length $h_{i,+}+h_{i,-}$ is sufficiently large.
A finite search limit allows the procedure to terminate when the cell is unbounded in a tested direction.

We require each coordinate interval to have length at least a prescribed threshold $h_{i,+}+h_{i,-} > L_{\min}$.
We retain the estimates for coordinates whose intervals satisfy this criterion and retry only the remaining coordinates from other points consistent with the same cell, using at most $T$ candidate starting points in total.
If, after these attempts, any coordinate still lacks a sufficiently long interval, we discard this cell for the current signature-recovery attempt.
This does not rule out recovering the same neuron from another pair of adjacent cells.
The length threshold is a numerical safeguard rather than a guarantee of a particular estimation error.

\paragraph{Rotating adjustment}
As a further refinement, we use directional probes to correct an initial
estimate of the row signature.
For two adjacent cells $C_0,C_1$ with a nonzero rank-one Jacobian difference,
the matrix that would be obtained in the absence of numerical errors has
the exact factorization
\begin{equation}
\label{eq:rotation-rank-one}
    \bm\Delta(C_0, C_1) = \bm J_{|C_0}-\bm J_{|C_1}
    = \lambda\bm a\bm n^{\intercal},
    \qquad \|\bm a\|=\|\bm n\|=1,\quad \lambda>0,
\end{equation}
where $\bm n$ is the true row-signature direction and $\bm a$ is the direction
of the corresponding change in the output derivative.
The preceding estimation procedure, followed by SVD, provides initial
estimates $\widehat{\lambda}$, $\widehat{\bm a}$, and
$\widehat{\bm n}$ of these quantities.
For a unit direction $\bm v$, the difference between the directional
derivatives in the two cells is
$\bm\Delta (C_0, C_1) \bm v=\lambda\bm a(\bm n^{\intercal}\bm v)$.
Thus, a direction orthogonal to $\widehat{\bm n}$ can reveal a residual
component of the true signature that the initial estimate misses.

To measure this difference, we choose a point $\bm x_j$ in each cell
$C_j$, $j\in\{0,1\}$, and nonnegative displacements
$h_{j,+}(\bm v),h_{j,-}(\bm v)$ such that
\[
    [\bm x_j-h_{j,-}(\bm v)\bm v,
     \bm x_j+h_{j,+}(\bm v)\bm v]\subseteq C_j,
    \qquad L_j(\bm v)=h_{j,+}(\bm v)+h_{j,-}(\bm v)>0.
\]
The centers and lengths may be chosen separately for each direction,
using the maximal-interval search described in the previous paragraph.
Define
\begin{align}
\label{eq:rotation-cell-probe}
    \bm D_j(\bm v)
    &=\frac{F(\bm x_j+h_{j,+}(\bm v)\bm v)
            -F(\bm x_j-h_{j,-}(\bm v)\bm v)}{L_j(\bm v)},\\
\label{eq:rotation-directional-jump}
    \bm d(\bm v)
    &=\bm D_0(\bm v)-\bm D_1(\bm v)
      =\lambda\bm a(\bm n^{\intercal}\bm v).
\end{align}
The last equality holds in exact arithmetic when both intervals remain
in their respective cells.
Dividing each cell's output increment by its own interval length allows
unequal and asymmetric intervals to be used.
If each endpoint evaluation in $C_j$ has error of norm at most $\eta_j$,
the evaluation-error contribution to the directional jump is bounded by
\[
    \frac{2\eta_0}{L_0(\bm v)}+\frac{2\eta_1}{L_1(\bm v)},
\]
before accounting for arithmetic rounding.
Thus, for fixed endpoint-error bounds, longer intervals improve this
bound, although a short or noisy interval on one side can dominate it.

We complete $\widehat{\bm n}$ to an orthonormal basis
$\{\widehat{\bm n},\bm t_1,\ldots,\bm t_{n_0-1}\}$ of the input space.
Probing along the $\bm t_i$ measures the components missed by
$\widehat{\bm n}$, while an additional probe along
$\widehat{\bm n}$ supplies a reference component.
Let $\widehat{\bm d}(\bm v)$ denote the numerical estimate of
\eqref{eq:rotation-directional-jump}.
Consider the ratios
\begin{equation}
\label{eq:rotation-coefficients}
    \widehat{r}_i
    =\frac{\widehat{\bm a}^{\intercal}\widehat{\bm d}(\bm t_i)}
           {\widehat{\bm a}^{\intercal}\widehat{\bm d}(\widehat{\bm n})},
    \qquad i=1,\ldots,n_0-1.
\end{equation}
If we had access to exact directional measurements $\bm d(\bm t_i) =\lambda\bm a(\bm n^{\intercal}\bm v)$, the common factor
$\lambda\widehat{\bm a}^{\intercal}\bm a$ would cancel, giving
$r_i=(\bm n^{\intercal}\bm t_i)/
(\bm n^{\intercal}\widehat{\bm n})$, provided both the common
factor and the reference component are nonzero.
This would yield the corrected direction
\begin{equation}
\label{eq:rotation-correction}
    \widehat{\bm n}_{\mathrm{corr}}
    =\frac{\widehat{\bm n}
            +\sum_{i=1}^{n_0-1}\widehat{r}_i\bm t_i}
           {\sqrt{1+\sum_{i=1}^{n_0-1}\widehat{r}_i^2}}.
\end{equation}
In exact arithmetic, this recovers $\bm n$ up to the usual global sign;
in finite precision, it can still give a candidate correction to the initial estimate.

Long intervals along the approximate tangent directions may improve
these measurements, but the cell-consistency checks remain heuristic.
When the initial estimates $\widehat{\bm n}$ and $\widehat{\bm a}$
are well aligned with their true directions, the exact reference signal
\[
    \widehat{\bm a}^{\intercal}\bm d(\widehat{\bm n})
    =\lambda(\widehat{\bm a}^{\intercal}\bm a)
             (\bm n^{\intercal}\widehat{\bm n})
\]
has magnitude close to $\lambda$.
Under this assumption, angular misalignment is not expected to make the
denominator in \eqref{eq:rotation-coefficients} small.
The reference signal may nevertheless be weak relative to measurement
uncertainty if the Jacobian jump itself is small.
Even when the reference signal is reliable, the small tangential
components needed for the correction may be obscured by numerical noise.

\color{black}

\subsection{Layer determination for two-side signatures}

When there is just one hidden layer, we are always in the case where two adjacent cells differ by a neuron in the last (and only) layer.
This means that, from different pairs of adjacent cells, we can collect several signatures referring to the same layer.
A naive heuristic is to assume that each neuron $\eta^{(\ell-1)}_1, \ldots, \eta^{(\ell-1)}_{n_{\ell-1}},$ has a non-null probability $p_1, \ldots, p_{n_{\ell-1}} > 0$ to switch and that $p = \min\{ p_1, \ldots, p_{n_{\ell-1}}\}$. 
The hardest signature to get will be the one corresponding to the neuron that is the least likely to switch.
The expectation for the number of attempts before observing the least likely signature is $1/p$ attempts.
 
In the presence of several hidden layers, we can collect column signatures that do not correspond to the column of the matrix $\bm W^{(\ell)}.$
In general, there is no way to check if two different cells differ in the last layer or in an intermediate one.
With a similar argument to \cite{carlini2020} and \cite{canales23}, we expect that signatures corresponding to the columns of the last matrix will have a higher frequency than signatures observed as a consequence of a switch in some intermediate state.
To justify this assumption, recall Eq.~\eqref{eq:Signature_as_rank1}.
For two linear cells $C_0, C_1$ that differ only in the activation of the neuron $\eta_{j}^{(i)}$ we have
$$
    \bm \Delta(C_0, C_1) = \bm L_{i,C_0} \bm E_{j,j} \bm R_{i,C_0} = \bm c_{*,j} \bm r_{j,*}^{\intercal} \in \mathbb{R}^{n_{\ell} \times n_0},
$$

Suppose there exist two other cells $C_2, C_3$ that differ in the activation of the same neuron $\eta_j^{(i)}$.
If $i=1$, then we have $\bm R_{1,C_0} = \bm W^{(1)} = \bm R_{1,C_2}$, which means that the matrices $\bm \Delta(C_0, C_1)$ and $\bm \Delta(C_2, C_3)$ will all share the same row $\bm w^{(1)}_{j,*}$ up to some multiplicative scalar.
That is, we would observe the same signature twice.
An identical argument can be made for the columns if the neuron that switches on both pairs $C_0, C_1$ and $C_2,C_3$ lies in the last hidden layer $\ell-1$.
In particular, let $\eta^{(\ell-1)}_j$ be that neuron, then we have $\bm L_{\ell-1,C_0} = \bm W^{(\ell)} = \bm L_{\ell-1,C_2}$, which means that the matrices $\bm \Delta(C_0, C_1)$ and $\bm \Delta(C_2, C_3)$ will all share the same column $\bm w^{(\ell)}_{*,j}$ up to some scalar.

In all the other cases, as there is no reason why the activation patterns relative to the layer $i+1$ to $\ell$ are the same for two different cells, we expect that in general $\bm R_{i,C_0} \neq \bm R_{i,C_2}$ and $\bm L_{i,C_0} \neq \bm L_{i,C_2}.$
Thus, while we find two independent transitions of the same neuron, we will observe two different column signatures.

A challenge for the whole process is that there is no guarantee that the frequency argument will let us attribute the signature to the correct layer.
In \cite{carlini2020}, it was observed that the preimages of critical hyperplanes relative to the nodes in the second and successive layers are hyperplanes bent by different activation patterns in the first layer.
This means that the signatures relative to the neurons in the first layer are always the same throughout the measurements, whereas the signatures relative to deeper nodes are ``distorted" in different ways.

Another interesting observation was made in \cite{liu2025}, noticing that, treating the weights of the row signatures as random variables, the variance inside row signatures coming from deeper layers is larger compared to the variance of the attacked layer.

\subsection{Limits of column signatures}

Working with Jacobians allows us to recover both row and column signatures.
A tempting idea is to use column signatures to reconstruct the last layer and go backward.
This would allow us to peel the neural network on both sides instead of being limited to working only from the first towards the last layer.

Unfortunately, this approach presents two important limitations on the recovery of the biases and the signs.

\paragraph{Bias recovery} 
Starting their attack from the first layer, Carlini et al. \cite{carlini2020} could determine a critical point $\hat{\bm x}$ for which $\bm w_i^{\intercal} \bm \hat{\bm x} + b_i = 0$. From this equation, it is easy to recover the value of the $i$-th component of the bias vector $\bm  b$ as $b_i = -\bm w_i^{\intercal} \bm \hat{\bm x}.$

In our case, once we collect all the column-signatures from $\bm W^{(\ell)},$ and consider different inputs $\bm x_i$ indexed by $i \in I,$ we will have the equation
\begin{equation}\label{eq:Wy+b=x}
    \bm W^{(\ell)} \bm h^{(\ell -1)}(\bm x_i) + \bm b^{(\ell)} = F(\bm x_i).
\end{equation}
where $\bm h^{(\ell -1)}(\bm x_i)$ is the output of the $(\ell - 1)$-th layer at input $\bm x_i.$
We have no information about the vector $\bm h^{(\ell -1)}(\bm x_i)$ besides that it is the output of a component-wise $\Relu$, meaning its coordinates are all positive or null.
We also do not know exactly $\bm W^{(\ell)}$ but we know a matrix $\hat{\bm W}^{(\ell)}$ whose columns $\hat{\bm w}_1, \ldots,\hat{\bm w}_m$ are the column signatures of the columns of $\bm W^{(\ell)}.$ 
The relation between the original weights $\bm W^{(\ell)}$ and the signature matrix is given by $\hat{\bm W}^{(\ell)} = \bm W^{(\ell)} \bm M$ where $\bm M = \bm P \bm D$ is the product between a permutation $\bm P$ and a diagonal matrix $\bm D.$
If there is a solution $\bm y_i = \bm h^{(\ell -1)}(\bm x_i)$ for Eq.~\eqref{eq:Wy+b=x}, then $\hat{\bm y}_i = \bm M^{-1} \bm y_i$ will be a solution of
\begin{equation}\label{eq:hat_Wy+b=x}
    \hat{\bm W}^{(\ell)} \hat{\bm y}_i + \bm b^{(\ell)} = F(\bm x_i).
\end{equation}
For each of these equations, the weight matrix $\hat{\bm W}^{(\ell)}$ and the outputs $F(\bm x_i)$ are known, but both $\bm b^{(\ell)}$ and $\hat{\bm y}_i$ are unknown.
From Eq.~\eqref{eq:hat_Wy+b=x}, it follows that $F(\bm x_i) \in \Colspan(\hat{\bm w}_1, \ldots, \hat{\bm w}_m, \bm b^{(\ell)})$ for all $i$; then $\bm b^{(\ell)} \in \cap_{i\in I} \Colspan(\hat{\bm w}_1, \ldots, \hat{\bm w}_m, F(\bm x_i)).$
When $\Rank(\hat{\bm W}^{(\ell)}) < n_\ell -1$ the spaces $\Colspan(\hat{\bm w}_1, \ldots, \hat{\bm w}_m, F(\bm x_i)) \subsetneq \R^{n_\ell}$ are strictly contained in $\R^{n_\ell}$ but, at best, we can determine $\bm b^{(\ell)}$ only up to an element in $\Colspan(\hat{\bm w}_1, \ldots, \hat{\bm w}_m)$ that is contained in each subspace of the form $\Colspan(\hat{\bm w}_1, \ldots, \hat{\bm w}_m, F(\bm x_i))$.

\paragraph{Sign recovery} 
Determining the correct sign for the column-signatures is also challenging.
A simple sufficient case in which it would be easier is if there is no bias (or if we could assume to know the bias from another source) and that the the matrix $\bm W^{(\ell)}$ has full column rank $\Rank(\bm W^{(\ell)}) = n_\ell$.
That is, if the equation
$$
    \bm W^{(\ell)} \bm y + \bm b^{(\ell)} = F(\bm x)
$$
admits a unique solution.

Suppose that we manage to recover and distinguish all the column-signatures of the matrix $\bm W^{(\ell)}$ and denote by $\hat{\bm W}^{(\ell)}$ the matrix whose columns are those signatures. 
To determine the correct sign of these columns, we can start by setting them positive (i.e., having the first non-zero component positive).
Observe that each component of $h^{(\ell -1)} (\bm x)$ is the output of a $\Relu,$ then the solution $\bm y$ must have all its components $\geq 0.$ 
If some components of the solution $\bm y$ we found are negative, it means we must multiply the corresponding column in $\hat{\bm W}$ by $-1$; in this way, we will force each component of the solution to be positive, while determining the correct sign of each signature.
If a coordinate of $\bm y$ is zero because the corresponding neuron is inactive, that observation provides no information about the sign of the associated column, and several inputs may therefore be required. Moreover, if $\bm W^{(\ell)}$ does not have full column rank, the solution $\bm y$ is not unique and non-negativity alone may not determine the column signs.

The scope of our contribution is limited to the signature-estimation primitive, which we evaluate experimentally in the following section, rather than a new end-to-end extraction attack. As in prior cryptanalytic extraction work, the vector estimator assumes access to raw multi-dimensional outputs. Although column signatures reveal additional structural information, they do not by themselves enable backward extraction, since ambiguities in bias, orientation, and layer attribution remain. The adaptive refinements additionally rely on heuristic
cell-consistency tests whose reliability decreases under low numerical precision.
\section{Experiments}
\label{sec:experiments}

We call the network under attack the oracle, a black box we query with an input and read a raw output vector from, with no access to weights. We evaluate the two-sided estimator of Section~\ref{sec:twoside} against Carlini et al.'s row-signature estimator~\cite{carlini2020}, isolating the estimation step from critical-point search. Every method receives the same estimated critical point $x^*$ and estimates its own output row and coordinate signs from queries alone, never from the oracle's true weights. 

\subsection{Setup}
\label{subsec:experiments-setup}

We attack single-hidden-layer ReLU networks $F(x) = W_2\,\mathrm{ReLU}(W_1 x + b_1) + b_2$ with $784$-dimensional input and hidden widths $h \in \{8, 32\}$, under three simulated oracle precisions (float64, float32, float16) applied to both the forward pass and every finite-difference step. We run every method against two weight sources: random Gaussian weights and weights after training on MNIST (reaching 92\% test accuracy). We report angular error $d_\mathrm{angle} = 1 - |\hat u \cdot u| / (\lVert \hat u \rVert \lVert u \rVert) \in [0,1]$ between the recovered and true row direction, the maximum coordinate-wise deviation $d_\infty$ between the normalized and sign-aligned parameter vectors, including the bias, and the number of oracle queries that each estimator spends past $x^*$. Each query counts as a single oracle input returning the full output vector $F(x)\in\mathbb{R}^{n_\ell}$, not one per scalar component.

We compare against two variants of Carlini et al.'s attack. Their original construction and an adaptive version of it, and evaluate three variants of our own attack.
\begin{description}
\item[C-band] Carlini et al.'s two-point safety-band construction: for each direction, a fixed offset is added before probing.
\item[C-band-adaptive] the probing interval is maximized, constrained to remain inside the same cell, in each direction to improve measurement accuracy.
\item[J-first-row] an ablation of J-vector-fixed that reads $w$ directly off row $0$ of the same fixed-step $\bm \Delta$ (Eq.~\eqref{eq:Signature_as_rank1}) instead of taking its leading right singular vector, isolating whether SVD's pooling across rows is actually earning its keep.
\item[J-vector-fixed] our method reads directly from $\bm \Delta$ (Eq.~\eqref{eq:Signature_as_rank1}) at a fixed probing interval $\delta$ (Eq.~\eqref{eq:partial_d}), with no refinement.
\item[J-vector-adaptive] the ``Adaptive steps'' refinement from Section~\ref{sec:twoside}: the probing interval $\delta$ is grown independently per side to the largest interval passing a local consistency check, before reading $\bm \Delta$.
\item[J-vector-rotate] the ``rotating adjustment'' refinement from Section~\ref{sec:twoside}: a further correction pass that probes along directions perpendicular to the current signature estimate, to recover the residual component of the true neuron signature missed by the initial estimate.
\end{description}

\subsection{Results}
\label{subsec:experiments-results}

Table~\ref{tab:carlini-comparison} reports the median angular error $d_\mathrm{angle}$ at each precision, averaged over both widths. We report results separately for random and MNIST-trained weights.

We also report the joint direction-and-bias error $d_\infty$, computed similarly to~\cite{carlini2020}. We first rescale $(w,b)$ and $(\hat w,\hat b)$ to the same unit-normal hyperplane form. The bias is divided by the same norm as its corresponding weight vector. We then sign-align the estimates using
$s=\mathrm{sign}(w\cdot\hat w)$
and compute the largest absolute coordinate-wise difference between the two augmented vectors. Thus, $d_\infty$ is an $L_\infty$ error, not an average over coordinates.

For each precision, the table reports the median $d_\infty$ over trials and the median query cost. Values below $10^{-15}$ indicate recovery that is exact up to the numerical precision.

Finally, we include J-first-row. Instead of extracting $w$ as the leading right singular vector, J-first-row reads it directly from row~0 of the same fixed-step $\bm{\Delta}$. This isolates the contribution of the SVD step.

\begin{table}[ht]
\centering
\small
\caption{Median angular error $d_\mathrm{angle}$, median joint direction-and-bias error $d_\infty$, and median query cost, random vs.\ trained (MNIST) weights, by simulated oracle precision, with trials from widths $\{8,32\}$ pooled before taking a single median ($n=90$ random / $90$ trained at float64/float32, $44$ random / $20$ trained at float16). The valid-transition rate, the fraction of attempted clean single-neuron crossings that were usable, was $100\%$/$99\%$ (random/trained) at float64, $98\%$/$97\%$ at float32, and only $2.4\%$/$1.1\%$ at float16. Both weight sources estimate their own row, sign, and starting point from queries alone.
}

\label{tab:carlini-comparison}
\resizebox{\linewidth}{!}{%
\begin{tabular}{lccc}
\toprule
Method & $d_\mathrm{angle}$ f64 (rand/trained) & $d_\mathrm{angle}$ f32 (rand/trained) & $d_\mathrm{angle}$ f16 (rand/trained) \\
\midrule
C-band (Carlini)            & $<10^{-15}$ / $<10^{-15}$ & $1.5\times 10^{-4}$ / $7.5\times 10^{-5}$ & $2.6\times 10^{-1}$ / $3.3\times 10^{-1}$ \\
C-band-adaptive (Carlini)   & $<10^{-15}$ / $<10^{-15}$ & $5.3\times 10^{-3}$ / $7.0\times 10^{-3}$ & $8.6\times 10^{-1}$ / $7.2\times 10^{-1}$ \\
J-first-row (ours)          & $<10^{-15}$ / $<10^{-15}$ & $7.1\times 10^{-4}$ / $4.0\times 10^{-5}$ & $4.1\times 10^{-1}$ / $3.4\times 10^{-1}$ \\
J-vector-fixed (ours)       & $<10^{-15}$ / $<10^{-15}$ & $5.4\times 10^{-5}$ / $2.2\times 10^{-6}$ & $4.9\times 10^{-2}$ / $3.8\times 10^{-1}$ \\
J-vector-adaptive (ours)       & $<10^{-15}$ / $<10^{-15}$ & $5.5\times 10^{-3}$ / $1.1\times 10^{-2}$ & $2.4\times 10^{-2}$ / $7.9\times 10^{-1}$ \\
J-vector-rotate (ours)      & $<10^{-15}$ / $<10^{-15}$ & $4.3\times 10^{-3}$ / $9.0\times 10^{-3}$ & $8.2\times 10^{-3}$ / $7.6\times 10^{-1}$ \\
\midrule
Method & $d_\infty$ f64 (rand/trained) & $d_\infty$ f32 (rand/trained) & $d_\infty$ f16 (rand/trained) \\
\midrule
C-band (Carlini)            & $1.5\times 10^{-10}$ / $1.1\times 10^{-10}$ & $7.5\times 10^{-3}$ / $4.1\times 10^{-3}$ & $2.1\times 10^{-1}$ / $2.2\times 10^{-1}$ \\
C-band-adaptive (Carlini)   & $1.6\times 10^{-10}$ / $1.0\times 10^{-10}$ & $4.5\times 10^{-2}$ / $5.6\times 10^{-2}$ & $2.0\times 10^{-1}$ / $2.6\times 10^{-1}$ \\
J-first-row (ours)          & $1.9\times 10^{-9}$ / $5.0\times 10^{-10}$ & $1.5\times 10^{-2}$ / $3.9\times 10^{-3}$ & $2.0\times 10^{-1}$ / $2.6\times 10^{-1}$ \\
J-vector-fixed (ours)       & $4.0\times 10^{-10}$ / $1.0\times 10^{-10}$ & $2.8\times 10^{-3}$ / $8.2\times 10^{-4}$ & $1.3\times 10^{-1}$ / $2.2\times 10^{-1}$ \\
J-vector-adaptive (ours)       & $4.6\times 10^{-15}$ / $3.1\times 10^{-15}$ & $4.0\times 10^{-2}$ / $4.9\times 10^{-2}$ & $6.4\times 10^{-2}$ / $2.5\times 10^{-1}$ \\
J-vector-rotate (ours)      & $3.7\times 10^{-15}$ / $1.9\times 10^{-15}$ & $3.8\times 10^{-2}$ / $3.9\times 10^{-2}$ & $3.5\times 10^{-2}$ / $3.0\times 10^{-1}$ \\
\midrule
Method & Queries f64 (rand/trained) & Queries f32 (rand/trained) & Queries f16 (rand/trained) \\
\midrule
C-band (Carlini)            & 6\,268 / 6\,268 & 6\,208 / 6\,226 & 5\,758 / 5\,692 \\
C-band-adaptive (Carlini)   & 18\,804 / 18\,804 & 18\,772 / 20\,428 & 27\,290 / 26\,651 \\
J-first-row (ours)          & 1\,570 / 1\,570 & 1\,570 / 1\,570 & 1\,570 / 1\,570 \\
J-vector-fixed (ours)       & 1\,570 / 1\,570 & 1\,570 / 1\,570 & 1\,570 / 1\,570 \\
J-vector-adaptive (ours)       & 55\,348 / 55\,445 & 41\,918 / 42\,462 & 14\,588 / 17\,412 \\
J-vector-rotate (ours)      & 201\,738 / 201\,510 & 146\,110 / 145\,050 & 52\,172 / 58\,888 \\
\bottomrule
\end{tabular}%
}
\end{table}
Increasing the probing interval benefits J-vector-adaptive and J-vector-rotate only when the local consistency check can reliably distinguish true cell boundaries from numerical fluctuations. In float64, numerical fluctuations are much smaller than neuron activation jumps, allowing the interval to grow to the true boundary without significant error. At lower precisions, however, numerical fluctuations can mask boundary crossings, causing interval growth to overshoot into a neighboring cell and contaminate $\bm \Delta$. J-vector-fixed avoids this issue by using a short, fixed interval.

We verified this against oracle ground truth. For every step accepted by (\ref{eq:chord-consistency}), we compared the endpoint's true activation pattern, computed from the target network's privileged weights, with that of the base point. We sampled $20$ coordinate directions, both signs, and both sides ($\bm J_-, \bm J_+$) per trial, using the population from Table~\ref{tab:carlini-comparison}. At float64, none of the $14,400$ accepted intervals crossed a cell boundary. At float32, $55\%$ and $44\%$ of accepted intervals crossed a boundary for random and trained weights, respectively. At float16, these rates were $15\%$ and $3.5\%$. This lower rate at float16 than at float32, is explained by the consistency threshold (below), which is more conservative at float16 and so causes the adaptive search to reject more intervals and terminate earlier, before they can overshoot. Thus, overshooting directly explains the degradation of interval growth at reduced precision.

The consistency threshold is adapted to the numerical precision of the oracle. Reducing this threshold makes the test more conservative and can limit the acceptance of intervals that cross a cell boundary, but it can also cause clean intervals to be discarded. Its selection therefore involves a trade-off between accepting contaminated intervals and rejecting valid ones.

At float64, all methods achieve essentially exact recovery, with $d_\mathrm{angle}<10^{-15}$ for all six estimators. At float32, J-vector-fixed gives the strongest accuracy on random weights, outperforming C-band-adaptive by nearly two orders of magnitude and C-band by approximately a factor of $3$. On trained weights, its advantage over C-band-adaptive exceeds three orders of magnitude, while requiring approximately $4$--$12\times$ fewer queries.

J-first-row further shows the benefit of pooling information across rows. Replacing the single-row estimate with SVD pooling across all rows of $\bm \Delta$ reduces $d_\mathrm{angle}$ by approximately $13\times$ and $18\times$ for random and trained weights at float32, respectively, at identical query cost. At float16, the improvement is approximately $8\times$ for random weights. The trained-weight case is the only exception, where J-first-row slightly outperforms J-vector-fixed ($3.4\times10^{-1}$ versus $3.8\times10^{-1}$).

At float16, refined estimators are highly sensitive to the weight source. J-vector-rotate outperforms both Carlini variants on random weights, whereas J-vector-adaptive and J-vector-rotate degrade to approximately $0.7$--$0.8$ error on trained weights. J-vector-fixed performs similarly to C-band and C-band-adaptive in both cases. The trained-weight sample is also smaller ($n=20$ versus $44$), reflecting the lower frequency of clean single-neuron crossings.

In query cost, J-vector-fixed consistently requires fewer queries than both Carlini variants. J-vector-adaptive and J-vector-rotate are generally more expensive than C-band and C-band-adaptive, with the exception of float16, where J-vector-adaptive uses fewer queries than C-band-adaptive.

\subsubsection{Bias recovery.} 
Recovering a neuron means recovering its bias along with its row direction, since together they place the critical hyperplane in input space. Table~\ref{tab:bias-comparison} reports the same comparison for the normalized bias error $d_\mathrm{bias}$. Throughout, we recover the normalized hyperplane offset $b/\lVert w\rVert$, not the physical parameter $b$, consistent with the $d_\infty$ metric above. Both Carlini variants recover the bias from a single output coordinate, $\hat b = -\hat w^\top x^*$. Our three methods instead solve the full-vector least-squares problem of Section~\ref{sec:twoside}, using every coordinate's crossing to jointly estimate it.

\begin{table}[ht]
\centering
\small
\caption{Median normalized bias error $d_\mathrm{bias}$, random vs.\ trained (MNIST) weights.}
\label{tab:bias-comparison}
\resizebox{\linewidth}{!}{%
\begin{tabular}{lccc}
\toprule
Method & $d_\mathrm{bias}$ f64 (rand/trained) & $d_\mathrm{bias}$ f32 (rand/trained) & $d_\mathrm{bias}$ f16 (rand/trained) \\
\midrule
C-band (Carlini)            & $1.5\times 10^{-10}$ / $1.1\times 10^{-10}$ & $6.8\times 10^{-3}$ / $3.8\times 10^{-3}$ & $2.0\times 10^{-1}$ / $2.0\times 10^{-1}$ \\
C-band-adaptive (Carlini)   & $1.6\times 10^{-10}$ / $1.0\times 10^{-10}$ & $2.5\times 10^{-2}$ / $3.2\times 10^{-2}$ & $1.8\times 10^{-1}$ / $2.7\times 10^{-1}$ \\
J-vector-fixed (ours)       & $3.6\times 10^{-10}$ / $1.0\times 10^{-10}$ & $2.6\times 10^{-3}$ / $8.2\times 10^{-4}$ & $1.3\times 10^{-1}$ / $2.2\times 10^{-1}$ \\
J-vector-adaptive (ours)       & $4.2\times 10^{-15}$ / $2.6\times 10^{-15}$ & $3.2\times 10^{-2}$ / $4.1\times 10^{-2}$ & $5.1\times 10^{-2}$ / $2.1\times 10^{-1}$ \\
J-vector-rotate (ours)      & $3.6\times 10^{-15}$ / $1.6\times 10^{-15}$ & $3.0\times 10^{-2}$ / $3.4\times 10^{-2}$ & $2.6\times 10^{-2}$ / $2.2\times 10^{-1}$ \\
\bottomrule
\end{tabular}%
}
\end{table}
Query cost matches Table~\ref{tab:carlini-comparison}, since bias is computed directly from the neuron weights and the queries already used to detect the boundary between two cells.

J-vector-adaptive and J-vector-rotate recover the bias much more accurately than the Carlini methods at float64. Their error is around $10^{-15}$, compared with about $10^{-10}$ for the Carlini methods. This improvement comes from estimating the boundary offset using all available output coordinates rather than relying on a single coordinate.
At float32 and float16, however, this advantage shrinks, since the directions (the neuron weights) computed by the two methods also degrade in precision.
\section{Conclusion}
\label{sec:conclusion}

 When a single neuron changes activation across two adjacent linear cells, the resulting Jacobian difference has rank one. Its right factor contains the row-side information exploited by previous extraction attacks, while its left factor provides complementary column-side information. By applying SVD to the full Jacobian difference, the information from all output components can be combined to obtain more accurate signature estimates. Our experiments show that this improves signature recovery under finite precision, particularly in the float32 regime, without increasing the number of oracle queries.

We also considered adaptive refinements based on enlarging the intervals used to estimate the Jacobians. Their main limitation in float32 and float16 was the reliability of the cell-membership test: accepting points outside the intended cell contaminates the estimated Jacobian difference. A natural direction for future work is therefore to develop more reliable membership tests under reduced numerical precision. The column-side information also exposes a new leakage channel; however, using it to extract layers beyond the last and combining it with forward extraction into a complete bidirectional procedure remain open problems.
\bibliographystyle{splncs04}
\bibliography{bibliography}

\end{document}